# Thermoelectric phase diagram of the SrTiO$_3$–SrNbO$_3$ solid solution system

Yuqiao Zhang[1], Bin Feng[2], Hiroyuki Hayashi[3], Tetsuya Tohei[2], Isao Tanaka[3], Yuichi Ikuhara[2] & Hiromichi Ohta[4,*]

[1]*Graduate School of Information Science and Technology, Hokkaido University, N14W9, Kita, Sapporo 060-0814, Japan*

[2]*Institute of Engineering Innovation, The University of Tokyo, 2-11-16 Yayoi, Bunkyo, Tokyo 113-8656, Japan*

[3]*Department of Materials Science and Engineering, Kyoto University Yoshida-Honmachi, Sakyo, Kyoto 606-8501, Japan*

[4]*Research Institute for Electronic Science, Hokkaido University, N20W10, Kita, Sapporo 001-0020, Japan*

[*]To whom all correspondence should be addressed: hiromichi.ohta@es.hokudai.ac.jp

**Thermoelectric energy conversion — the exploitation of the Seebeck effect to convert waste heat into electricity — has attracted an increasing amount of research attention for energy harvesting technology. Niobium-doped strontium titanate (SrTi$_{1-x}$Nb$_x$O$_3$) is one of the most promising thermoelectric material candidates, particularly as it poses a much lesser environmental risk in comparison to materials based on heavy metal elements. Two-dimensional electron confinement, *e.g.* through the formation of superlattices or two-dimensional electron gases, is recognized as an effective strategy to improve the thermoelectric performance of SrTi$_{1-x}$Nb$_x$O$_3$. Although electron confinement is closely related to**

the electronic structure, the fundamental electronic phase behavior of the SrTi$_{1-x}$Nb$_x$O$_3$ solid solution system has yet to be comprehensively investigated. Here, we present a thermoelectric phase diagram for the SrTi$_{1-x}$Nb$_x$O$_3$ (0.05 ≤ $x$ ≤ 1) solid solution system, which we derived from the characterization of epitaxial films. We observed two thermoelectric phase boundaries in the system, which originate from the step-like decrease in carrier effective mass at $x$ ~ 0.3, and from a local minimum in carrier relaxation time at $x$ ~ 0.5. The origins of these phase boundaries are considered to be related to isovalent/heterovalent B-site substitution: parabolic Ti 3d orbitals dominate electron conduction for compositions with $x$ < 0.3, whereas the Nb 4d orbital dominates when $x$ > 0.3. At $x$ ~ 0.5, a tetragonal distortion of the lattice, in which the B-site is composed of Ti$^{4+}$ and Nb$^{4+}$ ions, leads to the formation of tail-like impurity bands, which maximizes the electron scattering. These results provide a foundation for further research into improving the thermoelectric performance of SrTi$_{1-x}$Nb$_x$O$_3$.

## I. INTRODUCTION

Thermoelectric energy conversion, in which waste heat is transformed into electricity by the Seebeck effect, is attracting significant research attention as a potential energy harvesting technology [1]. Generally, the performance of thermoelectric materials is evaluated in terms of a dimensionless figure of merit, $ZT = S^2 \cdot \sigma \cdot T \cdot \kappa^{-1}$, where $ZT$ is the figure of merit, $S$ is the thermopower (Seebeck coefficient), $\sigma$ is the electrical conductivity, $\kappa$ is the thermal conductivity, and $T$ is the temperature. Although the $ZT$ values of several commercialized thermoelectric materials such as Bi$_2$Te$_3$ ($ZT$~1@400 K) and PbTe ($ZT$~0.8@600 K) are roughly 1 [2], these heavy metal–based materials present environmental risks due to the toxicity of their constituent elements and their chemical and thermal instability. Recently, transition metal oxides (TMOs) including

Na$_x$CoO$_2$ ($x \sim 0.75$) [3], Ca$_3$Co$_4$O$_9$ [4], and SrTiO$_3$ [5-7] have drawn a high volume of research as high-temperature thermoelectric power generation materials: this group of materials are considered to be particularly suitable for these applications due to their chemical and thermal robustness, as well as the their comparatively low environmental risk [8].

Among the available TMOs, electron-doped SrTiO$_3$ has been one of the most extensively studied materials for thermoelectric applications [9]. In 2001, Okuda *et al.* [5] synthesized Sr$_{1-x}$La$_x$TiO$_3$ ($0 \leq x \leq 0.1$) single crystals by the floating-zone method, and reported that they yielded a large power factor ($S^2 \cdot \sigma$) of 2.8–3.6 mW m$^{-1}$ K$^{-2}$ at room temperature. After this, Ohta *et al.* reported the carrier transport properties of Nb- and La-doped SrTiO$_3$ single crystals (carrier concentration, $n \sim 10^{20}$ cm$^{-3}$) at high temperatures ($\sim 1000$ K) to clarify the intrinsic thermoelectric properties of these materials [6]. Furthermore, the experimental discovery of unusually large thermopower outputs from superlattices and two-dimensional electron gases in SrTiO$_3$ [10,11] spurred substantial research efforts into SrTiO$_3$ superlattices [12] and heterostructures [13] for thermoelectric applications; for example, a superlattice composed of 1 unit cell (uc) SrTi$_{0.8}$Nb$_{0.2}$O$_3$ and 10 uc SrTiO$_3$ exhibited giant thermopower, most likely due to an electron confinement effect. Although electron confinement is strongly correlated with the electronic structure [14], a full understanding of the fundamental electronic phase behavior of the SrTi$_{1-x}$Nb$_x$O$_3$ solid solution system has not yet been developed.

In this study, we clarify the thermoelectric phase diagram for the full range of the SrTiO$_3$–SrNbO$_3$ solid solution system (hereafter referred to as the SrTi$_{1-x}$Nb$_x$O$_3$ ss). Although high-quality single crystals of SrTi$_{1-x}$Nb$_x$O$_3$ species with $x > 0.1$ are not available due to the low solubility limit of Nb in the lattice [15], epitaxial films of these material compositions can be fabricated by pulsed laser deposition (PLD) [16]. As summarized in Fig. 1, pure SrTiO$_3$ (space group $Pm\bar{3}m$, cubic perovskite structure, $a$ = 3.905 Å) is an insulator with a bandgap of 3.2 eV, in which the bottom of the conduction band is composed of triply degenerate, empty Ti 3d-$t_{2g}$ orbitals, while the top of the valence band is composed of fully occupied O 2p orbitals [17]. The valence state of Ti ions in crystalline SrTiO$_3$ is 4+ (Ti 3d$^0$). On the other hand, pure SrNbO$_3$ (space group $Pm\bar{3}m$, cubic perovskite structure, $a$ = 4.023 Å) is a metallic conductor [18,19], in which the valence state of the Nb ion is 4+ (Nb 4d$^1$). In between SrTiO$_3$ and SrNbO$_3$ in the SrTi$_{1-x}$Nb$_x$O$_3$ ss, there are two possible types of valence state change in the Ti and Nb ions, as shown in Fig. 1 (b) and (c): in the case of isovalent substitution [Fig. 1 (b)], mole fluction of Ti$^{4+}$ proportionally decreases with increasing Nb$^{4+}$ ($x$); on the other hand, heterovalent substitution, in which two Ti$^{4+}$ or Nb$^{4+}$ ions are substituted by adjacent (Ti$^{3+}$/Nb$^{5+}$) ions, can occur as shown in Fig. 1 (c). Based on these considerations, we focused on the valence state change of Ti and Nb ions in the SrTi$_{1-x}$Nb$_x$O$_3$ ss.

In this study, we fabricated high-quality epitaxial films covering the full range of the SrTi$_{1-x}$Nb$_x$O$_3$ ss, and analyzed their crystal lattice parameters, as well as their electrical properties such as carrier concentration, Hall mobility, and thermopower, from which the carrier effective masses and relaxation times were derived. During these experiments, we observed that two thermoelectric phase boundaries exist within the system, which originate from the step-like decrease in carrier effective mass at $x$ ~ 0.3, and from the minimum in carrier relaxation time at $x$ ~ 0.5. The origin of these phase boundaries were analyzed by considering cases of isovalent/heterovalent B-site substitution. Parabolic Ti 3d orbitals were found to dominate electron conduction for compositions with $x$ < 0.3, whereas the Nb 4d orbitals became more influential for compositions with $x$ > 0.3. At $x$ = 0.5, tetragonal distortion of the lattice, in which the B-site was composed of predominantly [$Ti^{4+}$/$Nb^{4+}$] ions, led to the formation of tail-like impurity bands, which maximized the electron scattering. The results obtained in this study may be used as a foundation for further work that seeks to improve the thermoelectric performance of $SrTi_{1-x}Nb_xO_3$.

**II. EXPERIMENTAL**

Approximately 100 nm thick $SrTi_{1-x}Nb_xO_3$ ($x$ = 0.05, 0.1, 0.2, 0.3, 0.4, 0.5, 0.55, 0.6, 0.7, 0.8, 0.9, and 1.0) epitaxial films were fabricated by PLD using dense ceramic disks of a $SrTiO_3$–$SrNbO_3$ mixture. We selected insulating (001) $LaAlO_3$ (pseudo-cubic perovskite, $a$ = 3.79 Å) as the substrate. Growth conditions were precisely controlled, with a substrate temperature of 850 °C, oxygen pressure of ~ $10^{-4}$ Pa, laser fluence of 0.5–1 J $cm^{-2}$ $pulse^{-1}$, yielding a growth rate of 0.3 pm $pulse^{-1}$. Details of the PLD growth process of $SrTi_{1-x}Nb_xO_3$ developed by our group have been published elsewhere [7,20,21].

Crystallographic analyses of the resultant films were performed by X-ray diffraction (XRD, Cu K$\alpha_1$, ATX-G, Rigaku) and scanning transmission electron microscopy (STEM). TEM samples were fabricated using a cryo ion slicer (IB-09060CIS, JEOL). The thin film was observed using scanning transmission electron microscope (STEM) (JEM-ARM200CF, JEOL Co. Ltd) operated at 200 keV. High-angle annular dark-field (HAADF) images were taken with detection angle of 68 – 280 mrad. The electron energy loss spectra (EELS) were acquired in STEM mode by an Enfinium spectrometer (Gatan Inc) with energy resolution of 1 eV.

$\sigma$, carrier concentration ($n$), and Hall mobility ($\mu_{Hall}$) were measured at room temperature by a conventional d.c. four-probe method, using an In–Ga alloy electrode with van der Pauw geometry. $S$ was measured at room temperature by creating a temperature gradient ($\Delta T$) of ~4 K across the film using two Peltier devices (while using two small thermocouples to monitor the actual temperatures of each end of the SrTi$_{1-x}$Nb$_x$O$_3$ films). The thermo-electromotive force ($\Delta V$) and $\Delta T$ were measured simultaneously, and $S$-values were obtained from the slope of the $\Delta V$–$\Delta T$ plots.

Band structures for SrTi$_{1-x}$Nb$_x$O$_3$ ($x$ = 0, 0.25, 0.5, 0.75, and 1) were calculated based on the projector augmented-wave (PAW) method [22], as implemented using VASP code [23]. For these calculations, we adopted Heyd-Scuseria-Ernzerhof (HSE) hybrid functionals [24], and a plane-wave cutoff energy of 550 eV and a 6×6×6 $k$-point mesh for cubic-perovskite were employed in the total-energy evaluation and geometry optimization for the unit cells of SrTiO$_3$ and SrNbO$_3$, while special quasi-random structures (SQSs) were used in the case of SrTi$_{1-x}$Nb$_x$O$_3$ ($x$ = 0.25, 0.5, and 0.75). The SQSs with eight Sr or Ti sites were constructed by optimizing the correlation functions of seven types of independent pairs using the CLUPAN code [25].

**III. RESULTS AND DISCUSSION**

Figure 2 (a) summarizes the X-ray reciprocal space mappings (RSMs) around the ($\bar{1}$03) diffraction spot of LaAlO$_3$ (overlaid). Intense diffraction spots from ($\bar{1}$03) SrTi$_{1-x}$Nb$_x$O$_3$ are seen together with those from the LaAlO$_3$ substrate, indicating that incoherent hetero-epitaxial growth of the target materials occurred for all $x$ compositions. The peak positions of the diffraction spots from each composition correspond well with the cubic line ($q_z/q_x$ = −3), indicating that no epitaxial strain was induced in the films. It should be noticed that a slight tetragonal distortion was observed in the $x$ = 0.4 ($c/a$ = 1.0057) and 0.5 ($c/a$ = 1.0050) samples. From the RSMs of the SrTi$_{1-x}$Nb$_x$O$_3$ films, we extracted their lattice parameters using the formula $a$ =

$(2\pi/q_x \cdot 2\pi/q_x \cdot 6\pi/q_z)^{1/3}$. Figure 2 (b) plots the lattice parameters of the SrTi$_{1-x}$Nb$_x$O$_3$ film as a function of $x$: we observed an M-shaped trend, along with a general increase in the lattice parameter with increasing $x$. In order to analyze the changes in lattice parameter, we calculated the average ionic radii in the crystal structure and used Shannon's ionic radii as a comparison [26]: Ti$^{4+}$ (60.5 pm), Ti$^{3+}$ (67.0 pm), Nb$^{4+}$ (68.0 pm), and Nb$^{5+}$ (64.0 pm). In the ranges of $0.05 \leq x \leq 0.3$ and $x \geq 0.6$, the observed lattice parameters closely followed the heterovalent substitution line, suggesting that two Ti$^{4+}$ or Nb$^{4+}$ ions are substituted by adjacent (Ti$^{3+}$/Nb$^{5+}$) ions [27]. On the other hand, at $x = 0.4$ and $0.5$, the observed lattice parameter corresponded well with the isovalent substitution line; moreover, at $x = 0.5$, the B-site occupation of [Ti$^{4+}$/Nb$^{4+}$] was almost 100%, as shown in Fig. 2 (c).

In order to further clarify the occupation of B-sites by [Ti$^{4+}$/Nb$^{4+}$], we analyzed the atomic arrangement and electronic structure of the $x = 0.5$ film by STEM and EELS. Figure 3 (a) shows a cross-sectional HAADF-STEM image of a SrTi$_{0.5}$Nb$_{0.5}$O$_3$ film. Periodical mismatch dislocations with an interval of ~ 8.5 nm were seen at the heterointerfaces. If the strain in the thin-film were fully relaxed by such misfit dislocations, it would be possible to calculate the spacing between dislocations ($d$) from $d = \mathbf{b}/\delta$, where $\mathbf{b}$ is the Burgers vector and $\delta$ is the lattice mismatch between thin-film and substrate [28]: using the lattice parameters obtained from XRD [$\delta = (q_{x\text{sub}} - q_{x\text{film}})/q_{x\text{film}} = +0.0435$], a dislocation spacing of 8.7 nm was estimated, suggesting that the dislocations do fully relax the strain in the film.

Although superspots originating from (111) diffraction are often observed in AB$_{0.5}$B'$_{0.5}$O$_3$ compositions that crystallize in B-site–ordered double perovskite structures [29], these were not observed in the SrTi$_{0.5}$Nb$_{0.5}$O$_3$ film [Fig. 3 (b)], most likely due to the slight tetragonal distortion of the crystal structure. Figure 3 (c) shows the EELS spectra acquired around the Ti $L$ (c) and O $K$ edges (d); the reported EELS spectra of $Ti^{3+}/Ti^{4+}$ [30] and $Nb^{4+}/Nb^{5+}$ [31] are also plotted for comparison. In the Ti $L$ edge spectrum (c), $t_{2g}$ and $e_g$ peak splitting was clearly observed for Ti $L_3$, indicating that the dominant valence state of Ti is 4+. In the O $K$ edge spectra (d), two intense peaks (assigned as A and B) were clearly seen, with the intensity of peak B being higher than A, which was noted to be a characteristic feature of $Nb^{4+}$ in a previous study [31]; the peak intensity ratio A/B was calculated to be 0.66, which roughly corresponds with the $Nb^{4+}$ spectrum (0.66). From these results, we conclude that isovalent substitution of $Ti^{4+}/Nb^{4+}$ dominates in the $x = 0.5$ composition.

We then measured $\sigma$, $n$, and $\mu_{Hall}$ by Hall effect measurements. Generally, $\sigma$ was observed to increase with increasing $x$, as shown in Fig. 4 (a); additionally, it was noted that there was a sharp increase in $\sigma$ between $0.5 < x < 0.6$, suggesting the existence of an electronic phase boundary at $x \sim 0.5$. $n$ was also measured to increase with increasing $x$, which is almost correspond with the nominal value ($n=x$ in a unit cell) [Fig. 4 (b)]. For the mobility, as shown in Fig. 4 (c), $\mu_{Hall}$ for compositions with $x < 0.5$ remained almost constant at $\sim 6$ cm$^2$ V$^{-1}$ s$^{-1}$, which is consistent with other reported values [32,33]. As was observed for $\sigma$, a sharp increase in $\mu_{Hall}$ from $\sim 5$ cm$^2$ V$^{-1}$ s$^{-1}$ to $\sim 10$ cm$^2$ V$^{-1}$ s$^{-1}$ was observed between compositions with $x = 0.5$ and $x = 0.6$, after which the values steadily increased to $\sim 12$ cm$^2$ V$^{-1}$ s$^{-1}$ at $x = 1$ (*cf.* a value of $\sim 14$ cm$^2$ V$^{-1}$ s$^{-1}$ measured for this composition in a previous study [19]). From these results, we can conclude that the small $\mu_{Hall}$ dominates the electronic phase boundary at $x \sim 0.5$.

In order to further elucidate the origin of the electronic phase boundary, we took measurements of $S$ across the composition range: by measuring the values of both $n$ and *S*, we were then able to calculate the density of states effective mass ($m^*$), as outlined in equations (1)–(3) below. Not surprisingly, |*S*| was found to monotonically decrease with increasing *x*, as shown in Fig. 4 (d). We then calculated $m^*$ using the following relation between *n* and *S* [34],

$$S = -\frac{k_B}{e}\left(\frac{(r+2)F_{r+1}(\xi)}{(r+1)F_r(\xi)} - \xi\right) \quad (1)$$

where $k_B$, $\xi$, $r$, and $F_r$ are the Boltzmann constant, chemical potential, scattering parameter of relaxation time, and Fermi integral, respectively. $F_r$ is given by,

$$F_r(\xi) = \int_0^\infty \frac{x^r}{1+e^{x-\xi}}dx \quad (2)$$

and *n* by,

$$n_- = 4\pi\left(\frac{2m^*k_BT}{h^2}\right)^{3/2} F_{1/2}(\xi) \quad (3)$$

where *h* and *T* are the Planck constant and absolute temperature, respectively. Using the equations (1)–(3), we obtained values of $m^*$, which are presented in Fig. 4 (e): a step-like decrease of $m^*$ was observed with increasing *x* value. $m^*$ for compositions with *x* < 0.3 was calculated to be ~ 1.1 $m_0$, which corresponds well with the values reported by Okuda *et al.*[5], while for compositions with *x* > 0.3, $m^*$ was ~ 0.7 $m_0$. At *x* = 1 (SrNbO$_3$), $m^*$ further decreased to 0.5 $m_0$, agreeing well with values calculated in another study using spin-polarized DFT calculations (0.4 $m_0$) [35]. We speculate that there is a percolation threshold for compositions with *x* ~ 0.3, and that parabolic Ti 3d orbitals dominate electron conduction when *x* < 0.3, whereas the Nb 4d orbitals determine electron conduction when *x* > 0.3. Finally, the relaxation time, *τ*, was extracted from the $\mu_{Hall}$ and $m^*$ values (= e·*τ*·$m^{*-1}$): these results are shown in Fig. 4 (f). The *τ* values in the range 0.05 ≤ *x* ≤ 0.2 are ~ 4 fs, followed by a drop, reaching a minimum value of ~ 2 fs at *x* = 0.5. After *x* = 0.5, *τ* initially increases sharply to 3.5 fs at *x* = 0.6, with a much more gradual increase with *x* composition in the range 0.6 ≤ *x* ≤ 0.9. Lastly, there was a slight drop in *τ* value to 3.2 fs for the *x* = 1 composition. The trend's trough shape, in particular, the sharp increase of *τ* from ~ 2 to ~ 3.5 fs in the range 0.5 < *x* < 0.6, clearly indicates that $\tau$ dominates the electronic phase boundary at $x \sim 0.5$. Yamamoto *et al.* reported that $\tau$ of the SrTi$_{0.8}$Nb$_{0.2}$O$_3$ epitaxial films was significantly reduced by Sr-site substitution with Ca or Ba, due to the occurrence of tetragonal lattice distortion [20]: since slight tetragonal distortion was observed in the $x = 0.4$ and 0.5 films, as shown in Fig. 2 (a), we speculate that the tetragonal distortion also minimizes the $\tau$ in this set of thin-films.

Then, we plotted the thermoelectric power factor ($S^2 \cdot \sigma$) of the SrTi$_{1-x}$Nb$_x$O$_3$ ss together with values reported in previous studies (Fig. 5). In the region around $x \sim 0.05$ where the observed data (red circles / red line) overlap with the reported values from [10] (white circles / grey line), the two data sets match well. There are two local maxima in $S^2 \cdot \sigma$ — firstly at $x \sim 0.05$ (~ 2.5 mW m$^{-1}$ K$^{-2}$), and secondly at ~ 0.55 (~ 0.9 mW m$^{-1}$ K$^{-2}$). We also observed two thermoelectric phase boundaries in the system, occurring in the regions around $x \sim 0.3$ and $x \sim 0.5$, as shown in the inset. Since $S$ is strongly correlated with $m^*$, the sharp decrease in $m^*$ at $x \sim 0.3$ [Fig. 4 (e)] accounts for the concomitant reduction in $S$. In addition, the small value of $\tau$ at $x \sim 0.5$ (where $\tau$ was measured to reach a minimum) contributed greatly to the observed reduction in $\sigma$. In conclusion, in order to maximize the thermoelectric performance of the SrTi$_{1-x}$Nb$_x$O$_3$ ss, the use of compositions with $x \leq 0.2$ would be the most suitable.

Finally, we calculated the band structures for SrTi$_{1-x}$Nb$_x$O$_3$ ($x = 0, 0.25, 0.5, 0.75$, and 1) in order to elucidate the origin of the two electronic phase boundaries using 2×2×2 supercell model structures. Figure 6 summarizes the resultant total and orbital-projected density of states (DOS) of the SrTi$_{1-x}$Nb$_x$O$_3$ compositions around the bandgap. The energy origin was set at the Fermi level, $E_F$ (dotted line). $E_F$ was found to increase gradually with the increase in $x$. When $x > 0$, the $E_F$ was located higher energy side of the conduction band minimum (Ti 3d – Nb 4d hybridized orbital), indicating that the compositions in this range are degenerate semiconductors (or metals). The similar band structure between $x = 0$ and $x = 0.25$ compositions is consistent with the similar $m^*$ and $\tau$ values recorded experimentally from films with $x < 0.3$. Meanwhile, both $x = 0.5$ and $x = 0.75$ SrTi$_{1-x}$Nb$_x$O$_3$ were found to possess a tail-like feature in the DOS below the $E_F$. The major component of the tail at $x = 0.75$ was the Ti 3d, whereas at $x = 0.5$, Ti 3d and Nb 4d orbitals both made comparable contributions to the observed feature. The former corroborates the suggestion from the experimental results that Ti$^{3+}$ ions are present [Figs. 2 (b) and 2 (c)]. However, the latter is completely different from the experimentally obtained fact that isovalent substitution of Ti$^{4+}$/Nb$^{4+}$ dominates in the $x = 0.5$ composition. As mentioned above, we did not observe any superstructure such as B-site ordered Sr$_2$TiNbO$_6$. Although we used special quasi-random structures (SQSs) in the case of the band calculations of SrTi$_{1-x}$Nb$_x$O$_3$ ($x = 0.25, 0.5, 0.75$), B-site ordering still remained due to the finite cell size ($2\times2\times2$ supercell of cubic perovskite). This should be the main reason that the calculation result of $x = 0.5$ does not reflect the experimentally obtained solid solution crystal. Moreover the characteristic decrease in $\tau$ measured for compositions around $x = 0.5$ (and was not observed when $x > 0.6$) may imply that the electrons from the Nb 4d orbitals in the tail-like DOS do not contribute to the electrical conductivity of the SrTi$_{1-x}$Nb$_x$O$_3$ ss.

**IV. SUMMARY**

In summary, we have clarified the thermoelectric phase diagram for the SrTi$_{1-x}$Nb$_x$O$_3$ ($0.05 \leq x \leq 1$) solid solution system through characterization of epitaxial thin-films. We observed two thermoelectric phase boundaries in the system, which originate from the combination of a step-like decrease in carrier effective mass at $x \sim 0.3$ with the local minimum carrier relaxation time at $x \sim 0.5$. The origins of these electronic phase boundaries were analyzed in the context of isovalent/heterovalent B-site substitution. Parabola-shaped Ti 3d orbitals dominate electron conduction for compositions $0 < x < 0.3$, whereas the Nb 4d orbitals dominate when $x > 0.3$. At $x \sim 0.5$, a tetragonal distortion of the lattice occurs, in which the B-site is composed of predominantly [$Ti^{4+}/Nb^{4+}$] ions, leading to the formation of tail-like impurity bands in the density of states, which maximizes electron scattering.

These findings may prove useful in further improving and optimizing the thermoelectric performance of $SrTi_{1-x}Nb_xO_3$. Moreover, the analyses of the B-site ion valence states in this study provide valuable information that could be used in the design of other functional materials based on transition metal oxides.

**ACKNOWLEDGEMENTS**

This research was supported by Grants-in-Aid for Scientific Research on Innovative Areas "Nano Informatics" (25106003, 25106005, and 25106007) from the Japan Society for the Promotion of Science (JSPS). H.O. was supported by Grants-in-Aid for Scientific Research A (17H01314) and B (26287064) from the JSPS. Y.Z. thanks the China Scholarship Council (CSC) for a scholarship to study in Japan. This work was also supported in part by the Network Joint Research Center for Materials and Devices.

**References**

[1] D.M. Rowe, *CRC Handbook of Thermoelectrics*. (CRC Press, 1995); H. Julian Goldsmid, *Introduction to Thermoelectricity*. (Springer, 2010).

[2] T. M. Tritt and M. A. Subramanian, MRS Bull. **31**, 188 (2006); G. Jeffrey Snyder and Eric S. Toberer, Nature Mater. **7**, 105 (2008).

[3] I. Terasaki, Y. Sasago, and K. Uchinokura, Phys. Rev. B **56**, 12685 (1997); M. Lee, L. Viciu, L. Li, Y. Y. Wang, M. L. Foo, S. Watauchi, R. A. Pascal, R. J. Cava, and N. P. Ong, Nature Mater. **5**, 537 (2006).

4   A. C. Masset, C. Michel, A. Maignan, M. Hervieu, O. Toulemonde, F. Studer, B. Raveau, and J. Hejtmanek, Phys. Rev. B **62**, 166 (2000);   S. W. Li, R. Funahashi, I. Matsubara, K. Ueno, and H. Yamada, J. Mater. Chem. **9**, 1659 (1999).

5   T. Okuda, K. Nakanishi, S. Miyasaka, and Y. Tokura, Phys. Rev. B **63**, 113104 (2001).

6   S. Ohta, T. Nomura, H. Ohta, and K. Koumoto, J. Appl. Phys. **97**, 034106 (2005).

7   S. Ohta, T. Nomura, H. Ohta, M. Hirano, H. Hosono, and K. Koumoto, Appl. Phys. Lett. **87**, 092108 (2005).

8   K. Koumoto, R. Funahashi, E. Guilmeau, Y. Miyazaki, A. Weidenkaff, Y. F. Wang, and C. L. Wan, J. Am. Ceram. Soc. **96**, 1 (2013).

9   Hiromichi Ohta, Kenji Sugiura, and Kunihito Koumoto, Inorg. Chem. **47**, 8429 (2008);   J. W. Fergus, J. Eur. Ceram. Soc. **32**, 525 (2012).

10  Hiromichi Ohta, Sungwng Kim, Yorika Mune, Teruyasu Mizoguchi, Kenji Nomura, Shingo Ohta, Takashi Nomura, Yuki Nakanishi, Yuichi Ikuhara, Masahiro Hirano, Hideo Hosono, and Kunihito Koumoto, Nature Mater. **6**, 129 (2007).

11  H. Ohta, T. Mizuno, S. J. Zheng, T. Kato, Y. Ikuhara, K. Abe, H. Kumomi, K. Nomura, and H. Hosono, Adv. Mater. **24**, 740 (2012).

12  W. S. Choi, H. Ohta, and H. N. Lee, Adv. Mater. **26**, 6701 (2014);   P. Delugas, A. Filippetti, M. J. Verstraete, I. Pallecchi, D. Marre, and V. Fiorentini, Phys. Rev. B **88**, 045310 (2013).

13  T. A. Cain, S. Lee, P. Moetakef, L. Balents, S. Stemmer, and S. J. Allen, Appl. Phys. Lett. **100**, 161601 (2012);   I. Pallecchi, F. Telesio, D. F. Li, A. Fete, S. Gariglio, J. M. Triscone, A. Filippetti, P. Delugas, V. Fiorentini, and D. Marre, Nature Commun. **6**, 6678 (2015);   S. Shimizu, S. Ono, T. Hatano, Y. Iwasa, and Y. Tokura, Phys. Rev. B **92**, 165304 (2015).

14  L. D. Hicks and M. S. Dresselhaus, Phys. Rev. B **47**, 12727 (1993);   N. T. Hung, E. H. Hasdeo, A. R. T. Nugraha, M. S. Dresselhaus, and R. Saito, Phys. Rev. Lett. **117**, 036602 (2016).

15  C. Rodenbucher, M. Luysberg, A. Schwedt, V. Havel, F. Gunkel, J. Mayer, and R. Waser, Sci Rep-Uk **6**, 32250 (2016).

16  T. Tomio, H. Miki, H. Tabata, T. Kawai, and S. Kawai, J. Appl. Phys. **76**, 5886 (1994).

17  L. F. Mattheiss, Phys. Rev. B **6**, 4718 (1972).

| 18 | S. A. Turzhevsky, D. L. Novikov, V. A. Gubanov, and A. J. Freeman, Phys. Rev. B **50**, 3200 (1994); X. X. Xu, C. Randorn, P. Efstathiou, and J. T. S. Irvine, Nature Mater. **11**, 595 (2012). |
| 19 | D. Oka, Y. Hirose, S. Nakao, T. Fukumura, and T. Hasegawa, Phys. Rev. B **92**, 205102 (2015). |
| 20 | M. Yamamoto, H. Ohta, and K. Koumoto, Appl. Phys. Lett. **90**, 072101 (2007). |
| 21 | Y. Mune, H. Ohta, K. Koumoto, T. Mizoguchi, and Y. Ikuhara, Appl. Phys. Lett. **91**, 192105 (2007). |
| 22 | P. E. Blochl, Phys. Rev. B **50**, 17953 (1994). |
| 23 | G. Kresse and D. Joubert, Phys. Rev. B **59**, 1758 (1999); G. Kresse and J. Furthmuller, Phys. Rev. B **54**, 11169 (1996). |
| 24 | J. Heyd, G. E. Scuseria, and M. Ernzerhof, J. Chem. Phys. **124** (2006); A. V. Krukau, O. A. Vydrov, A. F. Izmaylov, and G. E. Scuseria, J. Chem. Phys. **125** (2006); J. Heyd, G. E. Scuseria, and M. Ernzerhof, J. Chem. Phys. **118**, 8207 (2003). |
| 25 | A. Seko, Y. Koyama, and I. Tanaka, Phys. Rev. B **80** (2009); A. Seko, (2007). |
| 26 | R. D. Shannon, Acta Crystallogr A **32**, 751 (1976). |
| 27 | Y. Ishida, R. Eguchi, M. Matsunami, K. Horiba, M. Taguchi, A. Chainani, Y. Senba, H. Ohashi, H. Ohta, and S. Shin, Phys. Rev. Lett. **100**, 056401 (2008). |
| 28 | Y. Ikuhara and P. Pirouz, Microsc. Res. Tech. **40**, 206 (1998). |
| 29 | P. K. Davies, J. Z. Tong, and T. Negas, J. Am. Ceram. Soc. **80**, 1727 (1997). |
| 30 | D. A. Muller, N. Nakagawa, A. Ohtomo, J. L. Grazul, and H. Y. Hwang, Nature **430**, 657 (2004). |
| 31 | C. L. Chen, Z. C. Wang, F. Lichtenberg, Y. Ikuhara, and J. G. Bednorz, Nano Lett. **15**, 6469 (2015). |
| 32 | O. N. Tufte and P. W. Chapman, Phys. Rev. **155**, 796 (1967). |
| 33 | H. P. R. Frederikse, W. R. Thurber, and W. R. Hosler, Phys. Rev. **134**, A442 (1964). |
| 34 | C. B. Vining, J. Appl. Phys. **69**, 331 (1991). |
| 35 | Y. T. Zhu, Y. Dai, K. R. Lai, Z. J. Li, and B. B. Huang, J Phys Chem C **117**, 5593 (2013). |

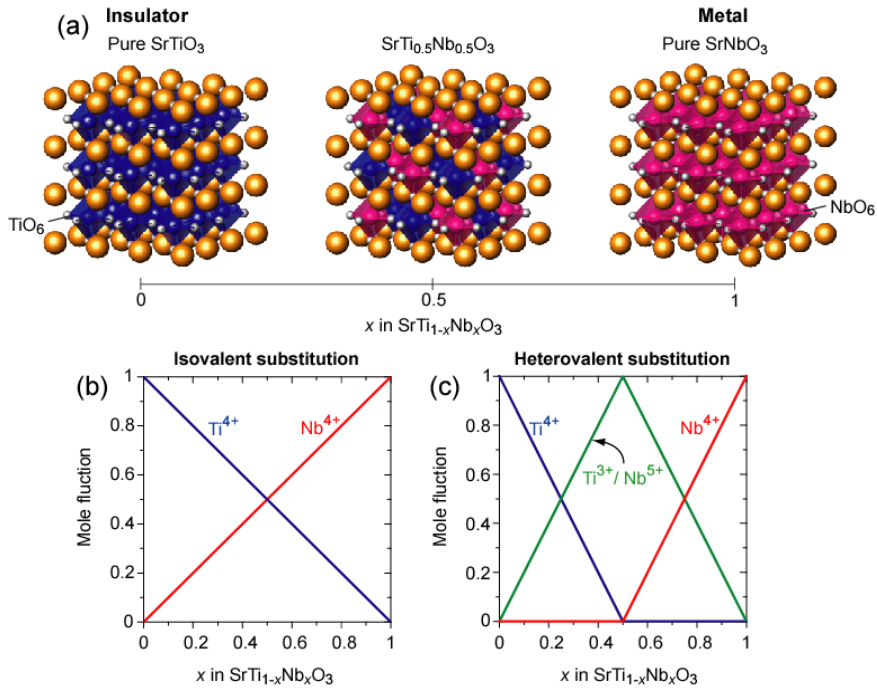

FIG. 1 (Color online)| Schematic crystal structure and possible valence state changes in the SrTiO$_3$–SrNbO$_3$ solid solution system. (a) Schematic crystal structure. Pure SrTiO$_3$ is an insulator with bandgap of 3.2 eV, in which the valence state of the Ti ions (blue, TiO$_6$) is 4+ (Ti 3d$^0$). On the other hand, pure SrNbO$_3$ is a metal, in which the valence state of the Nb ions (Red, NbO$_6$) is 4+ (Nb 4d$^1$). (b)–(c) Possible valence state changes of Ti and Nb ions in the SrTiO$_3$–SrNbO$_3$ solid solution system: (b) isovalent substitution, where Ti$^{4+}$ is substituted by Nb$^{4+}$; (c) heterovalent substitution, where two Ti$^{4+}$/Nb$^{4+}$ ions are substituted by adjacent Ti$^{3+}$/Nb$^{5+}$ ions.

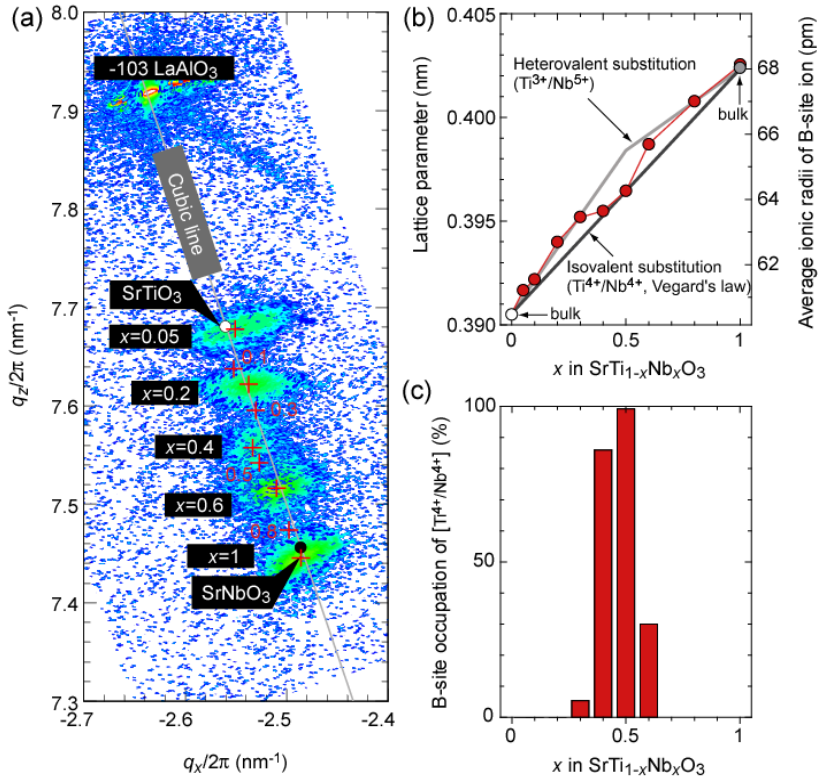

FIG. 2 (Color online)| Crystallographic characterization of the SrTi$_{1-x}$Nb$_x$O$_3$ epitaxial films on a (001) LaAlO$_3$ single crystal substrate. (a) X-ray reciprocal space mappings around the $(\bar{1}03)$ diffraction spot of the SrTi$_{1-x}$Nb$_x$O$_3$ epitaxial films. The location of the LaAlO$_3$ diffraction spot, $(q_x/2\pi, q_z/2\pi) = (-2.64, 7.92)$, corresponds with the pseudo-cubic lattice parameter of LaAlO$_3$ ($a = 0.379$ nm). Red symbols (+) indicate the peak positions of the SrTi$_{1-x}$Nb$_x$O$_3$ epitaxial films. (b) Changes in the lattice parameters of the SrTi$_{1-x}$Nb$_x$O$_3$ films (circles, left axis), with superimposed isovalent/heterovalent substitution lines (black line: isovalent substitution, gray line: heterovalent substitution, right axis), calculated using Shannon's ionic radii [26]. (c) Change in the B-site occupation by [Ti$^{4+}$/Nb$^{4+}$] derived from the data in (b).

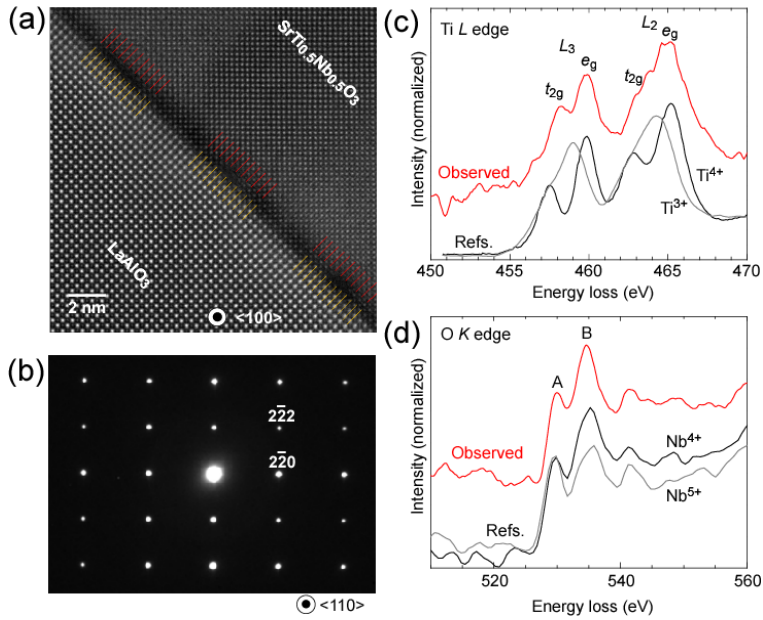

FIG. 3 (Color online)| Electron microscopy analyses of a SrTi$_{1-x}$Nb$_x$O$_3$ film with composition $x = 0.5$. (a) HAADF-STEM image acquired with the electron beam incident along the <100> direction. Periodic misfit dislocations (~ 8.5 nm interval) at the heterointerface are indicated by red lines. (b) Selected-area electron diffraction pattern acquired with the electron beam incident along the <110> direction. (c)–(d) EELS spectra acquired around the Ti $L$ edge (c) and O $K$ edge (d). EELS spectra for Ti$^{3+}$/Ti$^{4+}$ [30] and Nb$^{4+}$/Nb$^{5+}$ [31] from previous studies are also plotted for comparison.

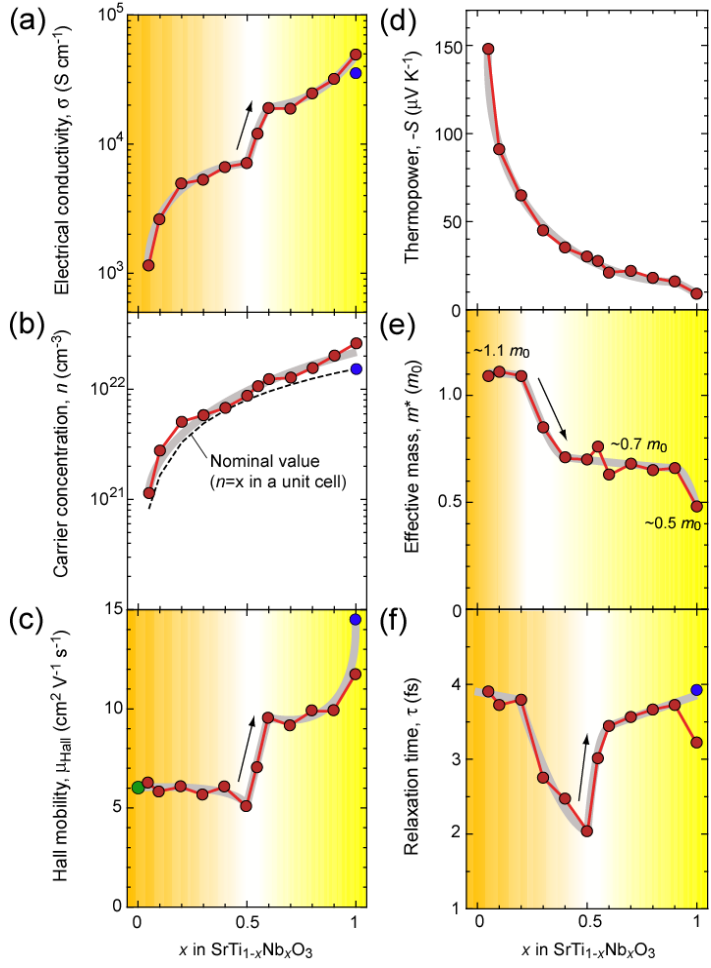

FIG. 4 (Color online)| Room temperature electron transport properties of the SrTi$_{1-x}$Nb$_x$O$_3$ epitaxial films: (a) carrier concentration ($n$), (b) electrical conductivity ($\sigma$), (c) Hall mobility ($\mu_{Hall}$), (d) thermopower ($S$), (e) effective mass ($m^*$), and (f) carrier relaxation time ($\tau$). Previously reported data for $\sigma$, $n$, and $\mu_{Hall}$ are plotted for comparison (blue circles indicate a SrNbO$_3$ film from [19]; green circles are for slightly Nb-doped SrTiO$_3$ [33]). The gray lines are drawn as a visual guide.

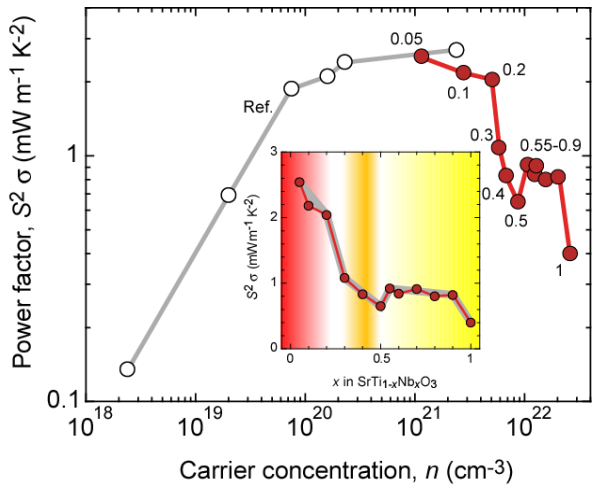

FIG. 5 (Color online)| Thermoelectric phase diagram for the $SrTiO_3$–$SrNbO_3$ solid solution system. The thermoelectric power factor ($S^2 \cdot \sigma$) of the $SrTiO_3$–$SrNbO_3$ solid solution system is plotted, alongside previously reported values [10]. The $x$ dependence of $S^2 \cdot \sigma$ is shown in the inset. The system's thermoelectric phase boundaries are clearly seen at $x \sim 0.3$ and $\sim 0.5$.

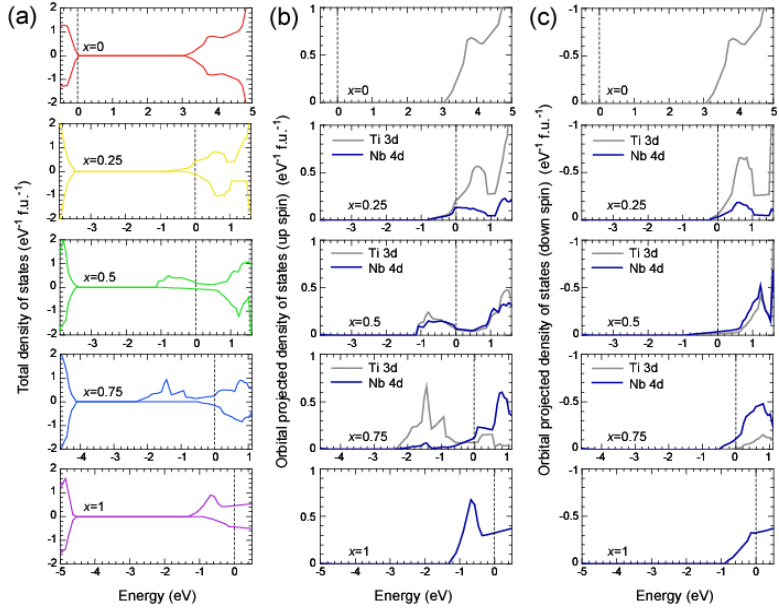

FIG. 6 (Color online)| Calculated density of states (DOS) for SrTi$_{1-x}$Nb$_x$O$_3$ ss ($x$ = 0, 0.25, 0.5, 0.75, and 1). (a) total DOS, (b)–(c) orbital-projected DOS. The energy origin was set at the Fermi level, $E_F$, indicated by the dotted line.